\documentclass[sigconf,authorversion,nonacm]{acmart}

\newcommand{\trigger}[1]{\textcolor{red}{{\bf Content Warning}: #1}}
\usepackage{subcaption}
\AtBeginDocument{%
  }

\begin{document}

\title{Characterizing Network Structure of Anti-Trans Actors on TikTok}


\author{Maxyn Leitner}
\affiliation{%
  \institution{University of Southern California Information Sciences Institute}
  \city{Los Angeles}
  \state{California}
  \country{USA}}
\email{maxynrl@isi.edu}

\author{Rebecca Dorn}
\affiliation{%
  \institution{University of Southern California Information Sciences Institute}
  \city{Los Angeles}
  \state{California}
  \country{USA}}
\email{rdorn@isi.edu}

\author{Fred Morstatter}
\affiliation{%
  \institution{University of Southern California Information Sciences Institute}
  \city{Los Angeles}
  \state{California}
  \country{USA}}
\email{fredmors@isi.edu}

\author{Kristina Lerman}
\affiliation{%
  \institution{University of Southern California Information Sciences Institute}
  \city{Los Angeles}
  \state{California}
  \country{USA}}
\email{lerman@isi.edu}


\begin{abstract}
\trigger{Trans-antagonistic Rhetoric and Terminology}
The recent proliferation of short form video social media sites such as TikTok has been effectively utilized for increased visibility, communication, and community connection amongst trans/nonbinary creators online.  However, these same platforms have also been exploited by right-wing actors targeting trans/nonbinary people, enabling such anti-trans actors to efficiently spread hate speech and propaganda.  Given these divergent groups, what are the differences in network structure between anti-trans and pro-trans communities on TikTok, and to what extent do they amplify the effects of anti-trans content?  In this paper, we collect a sample of TikTok videos containing pro and anti-trans content, and develop a taxonomy of trans related sentiment to enable the classification of content on TikTok, and ultimately analyze the reply network structures of pro-trans and anti-trans communities.  In order to accomplish this, we worked with hired expert data annotators from the trans/nonbinary community in order to generate a sample of highly accurately labeled data.  From this subset, we utilized a novel classification pipeline leveraging Retrieval-Augmented Generation (RAG) with annotated examples and taxonomy definitions to classify content into pro-trans, anti-trans, or neutral categories. We find that incorporating our taxonomy and its logics into our classification engine results in improved ability to differentiate trans related content, and that Results from network analysis indicate many interactions between posters of pro-trans and anti-trans content exist, further demonstrating targeting of trans individuals, and demonstrating the need for better content moderation tools.
\end{abstract}




\maketitle

\section{Introduction}
TikTok’s distinctive algorithmic and community dynamics make it a compelling case for studying the dual role of social media in fostering community-building and enabling targeted harassment, especially for marginalized groups like trans and nonbinary people. Platforms like TikTok allow trans/nonbinary creators to connect, share affirmations of identity, and disseminate essential resources about healthcare~\cite{aldridge_social_2024} and support systems~\cite{buss_transgender_2022}. Simultaneously, these platforms facilitate the proliferation of anti-trans rhetoric, weaponized by malign actors to spread hate speech and amplify harmful stereotypes \cite{srouji2022transgender}.

This work contributes to the understanding and modeling of these dynamics through three key innovations. First, we introduce a comprehensive taxonomy of sentiment toward trans and nonbinary communities, grounded in sociological and gender studies frameworks. This taxonomy addresses critical gaps in prior frameworks by including underexamined subcategories such as transmisogyny, anti-transmasculinity, and exorsexism, enabling a detailed and intersectional understanding of trans-related sentiment. This enables a more granular analysis of online discourse and its impact on marginalized communities.

Second, we develop a novel classification pipeline to identify pro-trans, anti-trans, and neutral content on TikTok, introducing a state-of-the-art classification pipeline that combines Retrieval-Augmented Generation (RAG) with curated annotated examples and taxonomy definitions, setting a new benchmark for classifying nuanced trans-related sentiment. Our RAG system is armed with two key resources: a curated set of annotated examples and the definitions from extant literature relating to our taxonomy. These resources not only guide the LLM’s contextual understanding but also significantly improve classification accuracy, especially in distinguishing nuanced pro-trans and anti-trans sentiments.

Finally, we apply our classification and taxonomy to analyze the reply network structures of pro-trans and anti-trans actors on TikTok. By comparing these network dynamics, we reveal the mechanisms by which anti-trans content proliferates and the structural characteristics that amplify its reach. This analysis informs potential strategies to mitigate online harms while empowering trans and nonbinary communities in digital spaces.

By combining sociological theory, novel computational techniques, and network analysis, this paper contributes a robust framework for studying the dynamics of marginalized communities and their adversaries in online ecosystems. These contributions advance the field of computational social science and offer actionable strategies for combating online harassment.


\section{Background and Related Work}
\subsection{Anti-Trans Rhetoric Online}
Due to its rapid proliferation and popularity across generational and political lines, TikTok makes for an incredibly effective social media space to analyze anti-trans and pro-trans content, as well as the ways in which they interact.  To begin, TikTok has served as a battle ground on which progressive views, particularly with respect to trans/non-binary people, have been scrutinized and exposed to attack from right wing actors \cite{hubrig_policing_2023,carless_when_2023}.  Right Wing content is incredibly popular on TikTok, and is often formatted in an attempt to appeal to the platform's younger user base \cite{weimann_research_2023, albertazzi_beyond_2024}. In concert with this, a combination of TikTok's content moderation approaches and recommendation algorithms often results in users being served increasingly right wing content, and anti-trans rhetoric serves as effective entry into that pipeline \cite{boucher2022down, little2021tiktok}. 

\subsection{Trans and Gender Diverse TikTok Users}
Structural forces such as content recommendation systems work at odds with trans/non-binary users, who report a lack of agency over being shown content that maligns their identities and convictions in ways that affect their beliefs about themselves \cite{ellen_simpson_for_2021, claudiu_gabriel_ionescu_are_2023}.  In the face of this opposition, trans/non-binary users persist and form folk theories in order to resist prevailing anti-trans narratives.  Where anti-trans actors are motivated to falsely report queer content as inappropriate as a way of training TikTok's content recommendation algorithms \cite{dawson_you_2024}, queer and trans users build and test theories of how to influence said algorithms in the other direction, towards greater visibility and acceptance of their identities \cite{nadia_karizat_algorithmic_2021}.  Transfeminine users have reported an awareness of how having their content promoted by the algorithm both opens them up to greater risk of attack from anti-trans actors, but also greater opportunity, moderating the former through strategies DeVito calls \textbf{algorithmic trapdoors} \cite{michael_ann_devito_how_2022}.  Given this landscape, TikTok is an important and illustrative platform on which to investigate the differences in network structure between anti-trans and pro-trans communities.   

\subsection{Perspectives of Trans Sentiment}
Previous work in classifying trans sentiment has mainly focused on the detection of anti-trans sentiment\cite{lu2022subtle, chakravarthi_how_2022}. However, these attempts fail to distinguish details in which trans people are targeted, and either presume or fail to analyze the identity of the source of anti-trans content, at the risk of visibilizing some forms of anti-trans sentiment over others. Given the painted heterogeneity of trans identities and experiences, previous taxonomy studies are limited both in the diversity of their subject pool and the specificity of anti-trans sentiment characterized \cite{Chang03072015}.


\section{Theoretical Grounding for a Trans\\ Sentiment Taxonomy}
We derive a taxonomy characterizing sentiment towards trans and gender diverse individuals in order to better understand and analyze online harms and social dynamics surrounding trans people. This work follows taxonomy design guidance \cite{kundisch2021update}, including specifying the taxonomy's goal and purpose, outlining qualitative techniques used, and evaluating the taxonomy's usefulness.
We begin with two particular articles which outline forms of Anti-Transgender sentiment that are frequently overlooked \cite{krell2017transmisogyny,nsambu_za_suekama_racial-class_2023}. Based on these sources, we continuously expand our search terms as more relevant articles are defined. Search terms used include ``cissexism", ``transandrophobia" and ``transmisogyny." See an extended list of search terms in the Datasheet.\footnote{https://github.com/maxynrl/TransTikTokDatasheet}



\noindent \emph{Relevance and Neutrality}
In societies constructed around cis binary gender as a norm, the consistency of anti-trans biases and the erasure of trans people from consideration can have significant impacts \cite{bauer_i_2009, blyth_death_2018}. Relatedly, research looking at trans-related sentiment in public opinion have struggled to find ``neutral" coverage due to increasing polarization along ideology \cite{jones_elite_2020}. Both of these realities are accurately inscribed by the minority stress model \cite{meyer2003prejudice} alongside the concept of ``cis-normativity'' \cite{serano_whipping_2016,kennedy2013cultural,enke2012education}. Combined with the prominent idea that all sides of an issue are equally morally valent (``both sideology"), equivocating views that cause differing degrees of material harm \cite{jost_bothsideology_2024}, we are called to rethink the concept of "neutrality" in this context.   In the scope of this work, we consider content ``Neutral" on trans issues when it is truly irrelevant to trans issues. 
Only content that is trans related is considered under either primary label Anti-Trans or Pro-Trans.  At a secondary tier, we consider a breakdown of specific kinds of Pro-Trans and Anti-Trans content via specific sub-categories.  These sub-categories are non-exclusive, meaning any particular message could be labeled with none, one or multiple secondary labels.

\noindent \emph{Anti-Trans}
In evaluating offensive content, situational context drastically changes harm \cite{zhou2023cobra}. We draw from seminal media literacy foundations, which posits that media sources have particular motivations and that the meaning of messages is altered by the consumer \cite{Jones-Jang_Mortensen_Liu_2021}. Accordingly, we incorporate context into our Anti-Trans sentiment label by considering two different aspects of the message in question: the message's target, and the source of the message.

For target, we value specificity in order to make visible targets of anti-trans sentiment that are currently understudied such as transmasculine \cite{avilalgbtq, banerjee2024trans}, non-binary \cite{pulice2024if, chan2019invisible}, and intersex people \cite{khanna2021invisibility}. Further, increased coverage helps to accurately capture specific ways that more hypervisible gender diverse people, namely transfeminine people \cite{gill-peterson_short_2024}, are targeted, and how this intersects with race (Blackness in particular) to form both heightened hypervisible precarity and ``hyperinvisibility'' \cite{ahern2024teaching, kennedy2023tight}.  For source, we consider the ways in which supposedly opposing groups collaborate in trans-antagonism \cite{schmidt2020conservatives, platero2023strange}, and also to accurately identify anti-trans content even when coming from members of the community.

With respect to the message's target, we thus consider the following sub labels: \textit{Transmisogyny}, \textit{Anti-Transmasculinity or Transandrophobia}, and \textit{Exorsexism}.
\textit{Transmisogyny}, coined by Julia Serano
\cite{serano_whipping_2016}, is a form of transantagonism targeting trans women and transfeminine people
The introductory work overlooks how class and race structure the impacts \cite{krell2017transmisogyny}. 
Alok Vaid Menon broadens the term, framing it within 
the ways in which transmisogyny targets all trans women and other transfeminine people, but is a part of a system of gender policing rooted in racism and anti-blackness resulting that can also affect cisgender people of color \cite{vaid2014trans}. 
It frames targets as ``threats" by sexualizing femininity as aggression 
\cite{gill-peterson_short_2024},
resulting in seemingly innocuous or positive phrases like ``protect women" carrying insidious, anti-trans implications as to who is considered a "woman", and from whom do they need ``protection"?   
\cite{turnbulldugarte_protect_2023, lee2014trans}. 
Sports exemplify how racialization informs transmisogyny in the construction of who is considered a proper ``girl" \cite{murib_dont_2022} 

\begin{figure*}[h]
    \centering    \includegraphics[width=.7\linewidth]{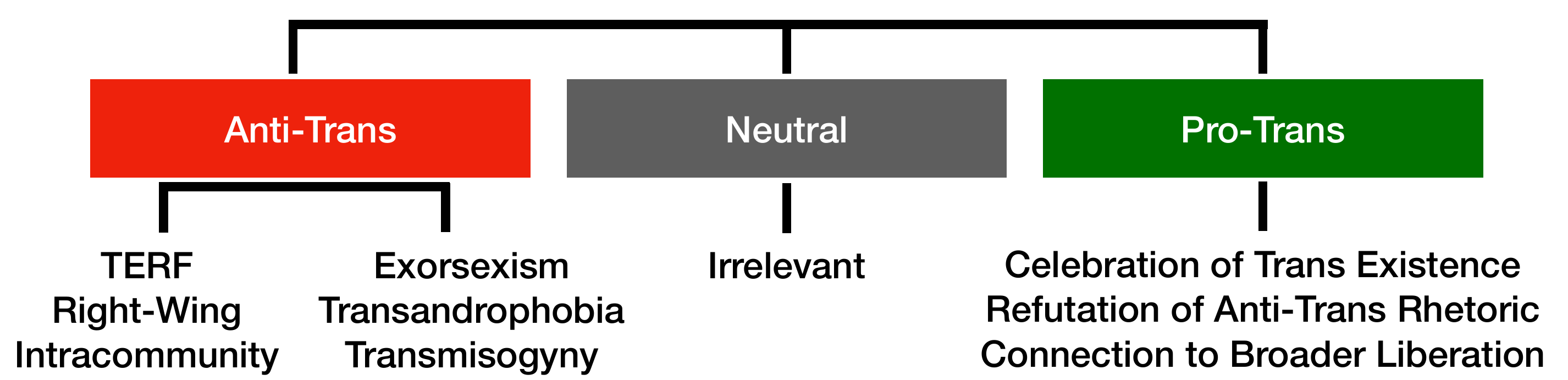}
    \caption{Trans Sentiment Taxonomy.}
    \label{fig:taxonomy}
    \vspace{-1em}
\end{figure*}

\textit{Anti-Transmasculinity} and \textit{Transandrophobia} are two terms that have gained popularity in naming the particular forms of anti-trans sentiment that target trans men and transmasculine people. Transmasculine people are often deliberately obscured in both society at large and trans discourse \cite{avilalgbtq}, as if to falsely state that they do not experience gender oppression in arenas that women, both cis and trans, are understood to \cite{bauer_i_2009, jones2020trans, jacobsen2022you, NORRIS2022100733}. Transmasculine people experience masculinity without the privilege of cis masculinity, often putting them in precarity with cis men \cite{abelson2014dangerous}, leaving them to navigate masculinity in the context of their gender oppression \cite{jeanes2021trans, rogers2019contrary}. When not completely invisibilized, transmasculinity is targeted by a kind of violent paternalism that considers them ``corrupted" into a ``trans cult", often by transfeminine people \cite{nsambu_za_suekama_racial-class_2023}. Anti-transmasculinity intersects with racism, particularly anti-blackness, to form Racialized Transmisandry, further compounding this portrayal of transmasculine people as ``gender traitors" with racist ideas about men of color being ``threatening"\cite{krell2017transmisogyny}. These contradictory rhetorics are are utilized to divide trans people against each other and deny transmasculine people the ability to name their oppression. 
 Because violence is considered masculine, and being subject to it is considered feminine, articulations of transmasculine oppression are incongruous with white, patriarchal gender logics \cite{betts2024black, nsambu_za_suekama_racial-class_2023}. Thus understandings of transmisogyny and anti-transmasculinity form a larger picture of anti-trans sentiment in concert with the concept of \textit{Exorsexism}. 
\textit{Exorsexism}, as coined by Tumblr user Vergess on May 5, 2016, refers to the idea that a person can only be exclusively male or female, invalidating non-binary and intersex existence entirely \cite{vergess_xor_2016}. The gender system we live under needs to sort people into two distinct boxes in order to maintain its heirarchy \cite{nsambu_za_suekama_racial-class_2023}, and our society's need to cling to the gender binary can be seen in the ways increasing acceptance of trans people comes conditioned on trans identities fitting neatly into these boxes \cite{worthen2021can}.  With respect to the message's target, we thus consider the following sub labels: \textit{Transmisogyny}, \textit{Anti-Transmasculinity/Transandrophobia}, and \textit{Exorsexism}.

When evaluating the source of the content we are evaluating, we consider three broad, overlapping categories: Trans Exclusionary Radical Feminists (aka TERFs), Right Wing, and Intracommunity.  TERFs base their ideology in aspects of second wave feminism, whereas Right Wing actors are based in religiously conservative right wing populism \cite{dickey2023transphobic, billard_gender-critical_2023, bassi2022introduction}. TERF and Right Wing approaches to Anti-Trans sentiment overlap in their demonization of trans people as a threat, but differ in the particulars. Both Right Wing and TERF anti-trans sentiment espouse bio-essentialist and gender essentialist beliefs, and frame cis women as threatened by trans people, transfeminine people especially. However, Right Wing trans-antagonism tends to emphasize ``Traditional Family Values'', claiming the existence of ``natural'' male and female gender roles, posing trans/non-binary people as threats to the family unit, children, and ultimately the nation-state \cite{nsambu_za_suekama_racial-class_2023, dickey2023transphobic}. 
 In contrast, TERFs bases misogyny as ``sex-based oppression'' rooted solely in bodily anatomy and reproductive organs, thus naming transfeminine people as ``invaders'' and transmasculine people as ``gender traitors'', both being threats to cis women \cite{dickey2023transphobic,billard_gender-critical_2023}. Analysis of Intracommunity Anti-Trans sentiment can draw upon any aforementioned avenues of trans-antagonism, and is important in naming ways in which community members leverage respectability politics and privilege to amplify harm against fellow community members\cite{finley2020respectability,Rosewarne04072023}. 

\noindent \emph{Pro-Trans}
We build our Pro-Trans taxonomy largely on approaches to pro-trans content we observed in our collected data while drawing on taxonomies related to online misogyny \cite{guest2021expert}, refined for trans-related sentiment \cite{chakravarthi2021datasetidentificationhomophobiatransophobia}.  First and foremost, we do not consider the source of the content as we do for anti-trans sentiment.  Given that large amounts of pro-trans content is created by trans/non-binary people, we want to avoid specifically identifying such users for their own privacy and safety.  Furthermore, even if such a determination relied solely on explicit self-reporting within the content, creators who at time of this paper identify themselves as cis may later realize that they are trans themselves.  Overall, any attempt at separating trans creators from cisgender allies is fraught and risks reinforcing cis-normativity, the normative assumption that all people are cisgender unless confirmed otherwise \cite{schmitt2019understanding, inmon2023imagining}.

We consider three categories that pro-trans content may or may not fall under: \textit{Celebration of Trans Existence}, \textit{Refutation of Anti-Trans Rhetoric}, and \textit{Connection to Broader Liberation}.  These categories were constructed after observing trends in pro-trans content on TikTok.  Once again, these categories are not mutually exclusive. The first of these categories, \textit{Celebration of Trans Existence}, applies to content that ranges from creators expressing joy at aspects of trans experiences, to simply identifying trans people identifying themselves as such while participating in hobbies, fan spaces, and other parts of online life.  This taps into the concept of \textit{Trans Joy}, a concept related to Chakravathi et al's ``Hope Speech'' \cite{chakravarthi2021datasetidentificationhomophobiatransophobia}. Celebrating the joy of being trans is an effective form of resistance against transantagonism, both by fighting the stigma associated with transness and by providing a vision of a joyful future for trans people \cite{westbrook2023transgender,lehva2023trans}.  Trans Joy is understudied in social science research \cite{doi:10.1177/15248399231152468, shuster_reducing_2022}, and including it in our taxonomy will bring into view pro-trans action that risks being invisibilized.

\textit{Refutation of Anti-Trans Rhetoric} is a straightforward category referring to content that directly engages with and refutes anti-trans content, and relates to Chakravarthi et al's ``Counter Speech'' \cite{chakravarthi2021datasetidentificationhomophobiatransophobia}. \textit{Connection to Broader Liberation} refers to content that explicitly makes a connection between transantagonism and other systems of oppression in hopes of building solidarity.  Examples of this can be seen in the connections between fat-antagonistic and trans-antagonistic body policing \cite{binder2023body}, related fights for bodily autonomy for both gender affirming and reproductive health care \cite{beck2024reproductive}, and the racism and anti-blackness that informs trans-antagonism \cite{vaid2014trans, krell2017transmisogyny, nsambu_za_suekama_racial-class_2023}.  With this, we have established the details of our taxonomy, as illustrated in Figure~\ref{fig:taxonomy}.

\subsection{The Trans Sentiment Taxonomy}
This taxonomy categorizes sentiment towards trans and gender diverse populations, making distinctions between subgroups and accounting for microaggression rhetoric. We present an overview of our taxonomy here.
At the top level of our taxonomy, we have three categories: Pro-Trans, Anti-Trans, and Neutral. Each category houses sub-categories.

\subsubsection{Pro-Trans}
\begin{itemize}
    \item \textit{Celebration of Trans Existence.} Explicit celebrations of trans life, and trans individuals identifying themselves while participating in joyful aspects of life (e.g. hobbies, communities).
    \item \textit{Refutation of Anti-Trans Rhetoric.} Engagement with Anti-Trans actors or rhetoric with the purpose of showing their flaws and/or the harm they cause to trans/non-binary people.
    \item \textit{Connection to Broader Liberation.} Content that draws parallels and connections between trans rights and other struggles for liberation, such as anti-racism, disability justice, reproductive freedom, and broader gender justice.
\end{itemize}
\subsubsection{Neutral}
\begin{itemize}
    \item \textit{Irrelevant.} Content that makes no explicit or implicit reference to trans/non-binary people, and has no implications for them specifically as trans/non-binary people.
\end{itemize}
\subsubsection{Anti-Trans}
\begin{itemize}
    \item \textit{Transmisogyny} Anti-Trans sentiment rooted policing claims to femininity and womanhood by those who fall outside the white, cisgender model of womanhood.  Particularly targets tranfeminine people as well as cisgender women of color, with this sentiment most sharply focused on Black transfeminine people via transmisogynoir.  
    \item \textit{Anti-Transmasculinity/Transandrophobia} Anti-Trans sentiment rooted in policing claims to masculinity and manhood, 
    \item \textit{Exorsexism} Anti-Trans sentiment rooted in the idea that one can only be male or female, not neither, and not both. Particularly targets nonbinary, intersex, and altersex people.
    \item \textit{TERF} Anti-Trans sentiment that comes from a Trans Exclusionary Radical Feminist source.  Typically appeals to a bioessentialist narrative of gender oppression, frames transness as rooted in "gender stereotypes" and a threat to the liberation and safety of cisgender women.
    \item \textit{Right Wing} Anti-Trans sentiment that comes from a Right Wing/Culturally Conservative Source.  Typically appeals to "traditional values" and frames transness as a threat to the family and national project.
    \item \textit{Intracommunity} Anti-Trans sentiment that comes from self-identified members of the trans/non-binary community.  Often appeals to respectability politics, and denigrating specific subgroups of the trans/nonbinary community in hopes of cisgender approval.
\end{itemize}

\section{Methods}
\subsection{Dataset}
We began data collection by identifying hashtags commonly used on TikTok that were associated with both pro-trans and anti-trans content by manually surveying prominent trans users and anti-trans figures on Twitter with over 1000 followers, and leveraging one author's domain expertise as a trans person following online trans discourse. This approach yielded a seed set of hashtags. We leveraged TikTok's research API~\cite{corso2024we} to gather videos that used these hashtags along with their associated comments. Next, we conducted a snowball sample, a process wherein an initial seed set of subjects are asked to recommend additional subjects, repeated for multiple rounds if needed to accrue subjects with similar characteristics to the seed set \cite{1521d86f-33df-3ece-8080-69edbcaba312, naderifar2017snowball}. In our implementation, we collected all hashtags that co-occurred in the content surfaced by our initial seed set, selected the hashtags relevant to trans/nonbinary identity, and then performed another round of data collection from the TikTok Research API by identifying other hashtags that commonly occur in these videos. With these additional hashtags, we collected more videos and comments.  Some examples of the anti-trans and pro-trans hashtags we searched over are listed in Table~\ref{tab:antihash}.

\begin{table}[t]
    \centering
    \begin{tabular}{|c|c|}
    \hline
 Anti-Trans & Pro-Trans \\ [0.5ex] 
         \hline
         \#nooneisborninthewrongbody & \#transgender\\ 
         \#saveoursinglesexspaces & \#transrights \\ 
         \textbf{\#genderideology} & \#transmen \\ 
         \textbf{\#whatisawoman} & \#nonbinaryawareness \\ 
         \#parentalrights & \#transmisogyny \\ 
         \textbf{\#ywnbaw} & \#protecttranskids \\ 
         \textbf{\#terftok} & \#transisbeautiful \\
         \textbf{\#savethetomboys} & \#tdov \\
         \#savewomenssports & \#tdor \\
         \textbf{\#adulthumanfemale} & \#nonbinaryvisibility\\
         
         \hline
    \end{tabular}
    \caption{Example Hashtags (seed set hashtags in bold).}
    \label{tab:antihash}
    \vspace{-3em}
\end{table}

Many of our anti-trans hashtags utilized euphemisms and dog whistles in order to avoid content moderation.  Explicit slurs denigrating trans people often yield no results, as TikTok refuses to serve content with these words. Of particular note is that different groups of trans people were targeted by particular hashtags.  For example, the hashtag ``\#ywnbaw,'' an acronym for ``You will never be a woman,'' targets trans women and other transfeminine people specifically, and was associated with a large number of videos on the platform.  However, the hashtag ``\#ywnbam'' (``You will never be a man'') returned relatively few results, as trans men and other transmasculine people were much more likely to be targeted with the hashtags ``\#genderconfusion'' and  ``\#savethetomboys.''

Additionally, some of the hashtags we crawled could lead to content that is either anti-trans, or completely unrelated, depending on the context.  For example, ``\#groomer'' could be falsely portraying trans people as inherently predatory, or it could be attached to dog grooming content.  Thus, we attempted to filter out results that contained hashtags indicating a context other than trans people.

In total, we gathered data on 59,860 videos as a result of searching on 41 anti-trans hashtags and 26 pro-trans hashtags. The resulting data entries included the username of the creator, the video's description, hashtags, time of creation, like count, comments and a link to the video.  We then downloaded each TikTok video via the python library Pyktok.\footnote{https://pypi.org/project/pyktok/} Pyktok is effective at downloading videos from the platform, but encounters several issues in video retrieval.  Reasonably, videos that have since been made private are not able to be downloaded.  TikTok allows posts that are simply a photo collage along with some background audio (usually music), which similarly cannot be downloaded.  Finally, with some probability Pyktok will fail to download public video content.  All three of these phenomena were reported in Pinto et al's work analyzing content on the platform \cite{pinto2024gettokgenaienrichedmultimodaltiktok}, and in our case affected 31.7\% of our collected data.  However, many of these samples still contained identifiable trans sentiment in the collected description, so we retained them in our dataset. 
\\
For all videos that we were able to download, we proceed to generate an automated transcription of the audio in the video. We utilized OpenAI's Whisper to transcribe audio for each sample \cite{radford2023robust}. For each of our samples, we construct a string composed of the audio transcription, followed by the written description associated with the TikTok, and associate it with that sample in our dataset.
\\
From here, we proceeded to parse the description of each video for interaction with other accounts. On TikTok, there are 4 kinds of interactions with other accounts that can occur through the creation of a video: Tags, Stitches, Duets, and Replies.  Tags refer to any other account that is simply ``tagged'' in the video description through the use of ``@.''  Multiple accounts can be tagged in the same video, as demonstrated in the example below:\\
\begin{quote}
    \texttt{the best of friends @reneerapp @dylanmulvaney\\ \#reneerapp \#reneerappsupremacy \#reneerappfan\\ \#dylanmulvaney \#everythingtoeveryone \#live}
\end{quote}

Stitches refer to content that consists of video from a creator appended to part of another creator's video.  This allows for a creator to make a public response to
another creator's content.  Meanwhile, duets allow a user to create a video consisting of their own content alongside another user's content, synced temporally.  Finally, Replies allow a user to create a video specifically in response to a comment on one of their previous videos.  These three interactions in particular can be used to form their own respective reply networks, as well as a larger reply network containing all three kinds of ``replies.''   For each video, we stored the usernames of any other users who were tagged, replied to, stitched with, or dueted with according to the kind of interaction.  At times, TikTok would resolve the tagged username to a non-unique display name, which allows for whitespace, and so we were unable to parse the identity of the users tagged in this way.     

\begin{figure*}[h!]
    \centering    \includegraphics[width=0.9\textwidth]{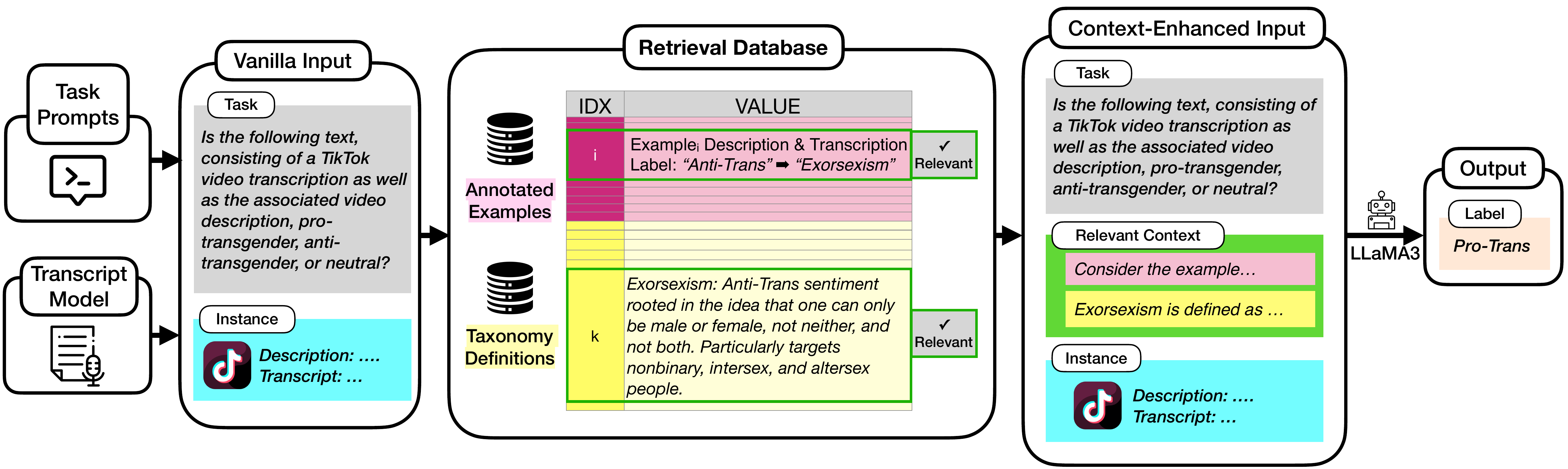}
    \caption{Pipeline detailing utilization of RAG (Retrieval Augmented Generation). Before the model is prompted, relevant annotated examples and taxonomy concepts are retrieved and added to the prompt. Enhancing the model with relevant examples and concepts leads to improved accuracy in pro-trans classification, lower false negatives in anti-trans classification, and improved accuracy overall.}
    \label{fig:pipeline}
    \vspace{-1em}
\end{figure*}

\subsection{Network Patterns}
Given that Duets and Stitches both involve reacting to another user's video with a video of one's own, we grouped both of these interactions together and created a reply network of these interactions in our dataset.  Likewise, we grouped Replies and Tagging together, as they involve reacting to either a user's comment or simply referring to them in a video description. We were able to visually identify both clusters centered around known anti-trans actors and clear pro-trans actors.  We aim to automatically identify users in these networks as pro-trans, anti-trans, or neutral based on their content, in order to assess differences in network structure between pro-trans and anti-trans communities.  For this purpose, we build a labeled dataset from our collected data using our taxonomy of trans related sentiment outlined earlier and visualized in Figure~\ref{fig:taxonomy}.  

\subsection{Annotation}
In order to accurately and ethically label our sample data, we follow recommendations in formulating our task, selecting our annotators, considering platform and infrastructure choices, as well as analyzing and evaluating our data \cite{denton_whose_2021}. 

\subsubsection{Annotation Set-Up}
We compile our labeled dataset by randomly selected 100 samples from each of the following: TikToks that contained \textbf{only Pro-Trans hashtags}, TikToks that contained \textbf{only Anti-Trans hashtags}, and TikToks that contained \textbf{both Pro-Trans and Anti-Trans hashtags}. This helps include as many samples belonging to each of our three main labels as possible.

To combat the subjectivity of our task, we make available a codebook containing sentences guiding how to classify content in accordance with our taxonomy. 
This helps normalize what terms we are using for the purposes of this labeling, how various edge cases should be classified, and what our standards for pro-trans, anti-trans, and irrelevant content are.
Example sentences in our codebook are provided in our datasheet.
         

\subsubsection{Annotator Selection}
Annotator selection impacts annotation accuracy, as identity and lived experience (or lack thereof) changes an annotator's ability to effectively recognize certain social stimuli \cite{scheuerman_datasets_2021}.  Furthermore, experience with anti-oppression frameworks is important to actively label oppressive content \cite{waseem-2016-racist}. 
Our annotation team consists of two members, both of which are internal researchers on the project. To account for the socio-cultural backgrounds and lived experiences of our annotators, annotators are members of the trans/non-binary community with experience in trans activism and frameworks for recognizing and countering anti-trans action. 
Our datasheet further discusses handling of annotation ethics, including how the people most effective at labeling anti-trans content will be those most affected by exposure to it. 

From here, we proceeded with the annotation of our 300 selected samples according to our taxonomy and codebook with our annotation team. One annotator labeled all 300 samples, while the other labeled a subset of 50 samples for validation.  Annotators yielded fairly high annotator agreement on this subset (Cohen Kappa 0.64), particularly for such a subjective task.

\begin{figure*}
    \includegraphics[width=\textwidth]{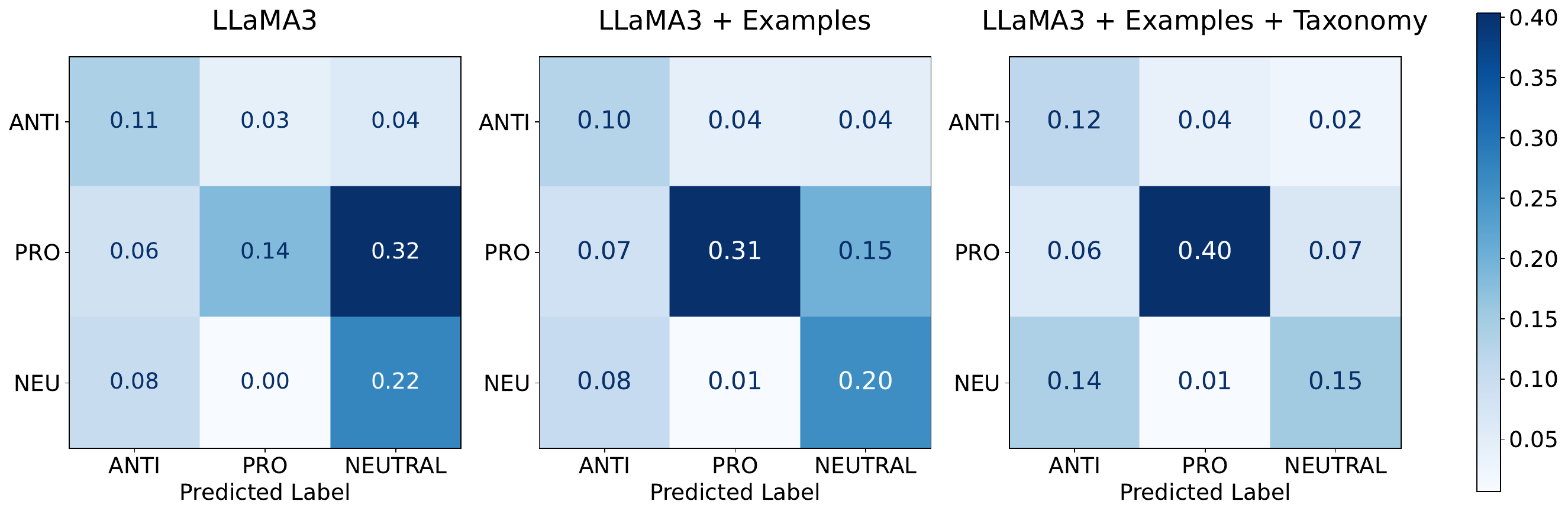}
    \caption{Classifier Confusion Matrices}
    \label{fig:zeroshot}
\end{figure*}

\subsection{Classification via LLM}
We build a trans sentiment classification model for TikTok, and experiment with improving performance by integrating our taxonomy with RAG (Retrieval Augmented Generation). We use LLaMA 3 as a base model, as it builds upon its previous iteration which has extensive reporting of safety protocols and training data specifically containing terms explicitly relating to gender and transgender identity in particular \cite{dorn_harmful_2024}.

 Processing raw videos is incredibly costly and difficult. Instead, model input includes an audio transcription from the Whisper API, \footnote{https://openai.com/index/whisper/} allowing a text representation of the audio. The Whisper API is a well-established tool for automated speech recognition, with demonstrated robustness across various domains and accents~\cite{radford2023robust}. Given the extent of its training data, we expect it to contain less bias than other transcription tools. To further mitigate bias, given potential bias in speech recognition \cite{johnson2023analysis}, we add the user's written video description.

\subsubsection{LLaMA3}
 We began by taking a zero shot approach, tasking the LLM with classifying each of our annotated examples simply using the pre-trained LLaMA 3 model.  We constructed 8 prompts tasking LLaMA 3 \cite{dubey2024llama} with labeling the constructed text sample as either anti-trans, pro-trans, or neutral.  For each sample, we tried zero shot classification with each of the 8 prompts, then ensembled the results.  We compared the ensembled labels to the ground truth labels for each sample, evaluating the model's recall, precision, F1, along with constructing a confusion matrix of the three classes We also kept track of which prompt performed the best over all samples, and it was that prompt we used for the following approaches.
 
\subsubsection{LLaMA3+RAG Examples}
In order to combat LLM hallucination and LLM difficulty to reason with out-of-text context, we supplement our zero-shot approach using the fine-tuning method Retrieval Augmented Generation (RAG)\cite{gao2024retrievalaugmentedgenerationlargelanguage}. RAG links an LLM input with an external database using an indexing procedure, in which the RAG system indexes the external database files by breaking them into smaller chunks, creating vector representations of these chunks for retrieval.  Data is then retrieved by computed relevancy to the text input in our case the transcription, description, and classification prompt, and provided to the LLM as additional context.  We utilize LlamaIndex as our RAG system, for its compatibility with the LLaMA 3 engine as well as overall performance \cite{zirnstein2023extended}. In LLaMA3+RAG Examples, we index annotated examples into two primary buckets: Anti-Trans and Pro-Trans. Examples are matched to inputs based on the cosine similarity of their representations, and anywhere from 0 to 3 examples can be matched.


\subsubsection{LLaMA3+RAG Examples+RAG Taxonomy}
We evaluate the efficacy of our taxonomy by bolstering the RAG system with our codebook.  The taxonomy is processed such that each definition is separately indexed, and appended to the RAG Examples database.  In this configuration, our RAG system is supplied with both example inputs and guiding principles to classify content samples.

\section{Results}
\subsection{Classification Results}

\subsubsection{LLaMA3}
The zero shot approach in Figure \ref{fig:zeroshot} exhibits lowest performance when distinguishing between Pro-Trans content and Neutral content. 

In Table \ref{tab:results}, we see an overall lean towards Neutral classification in the zero-shot approach, with a high false positive rate for that category (precision=0.37, recall=0.74). The inverse is true for Pro-Trans content with a high false negative rate (precision $=0.86$, recall $=0.28$). The zero-shot approach performs reasonably at classifying Anti-Trans content, with precision $> 0.4$ and recall $> 0.6$ for an F1 score $> 0.5$. When we examine recall by sublabel in Table \ref{tab:subrecall}, more details emerge. While this model struggles to recall Pro-Trans content in general, the effect is most pronounced with \textit{Celebration of Trans Existence} content (CEL). For Anti-Trans content, this model has low recall on transmisogynistic (TM), anti-transmasculine (ATM), and intracommunity (INTRA) content.

\subsubsection{LLaMA3+RAG Examples}
The addition of the RAG system populated with sample data led to an increased ability to differentiate Pro-Trans Content from Neutral Content, as seen in Figure \ref{fig:zeroshot} as well as the increased precision and F1 scores for both Pro-Trans and Neutral Categories in Table \ref{tab:results}.  Neutral false positive rate and Pro-Trans false negative rate have both been significantly reduced, without significant increases in Neutral false positive or Pro-Trans false negative rate.  Anti-Trans classification experiences a slight downturn in precision and recall, and thus also F1 score. Analyzing by sublabel, we see significant gains in recall for all Pro-Trans sublabels for this model in Table \ref{tab:subrecall}.  Meanwhile, recall for Anti-Trans content has decreased for al sublabels except for \textit{Intracommunity} (INTRA) and \textit{Transmisogyny} (TM).
\subsubsection{LLaMA3+RAG Examples+RAG Taxonomy} 
With the addition of the codebook to our RAG system, we see even further improvements leading to this model having the highest overall accuracy. This system excels at identifying Pro-Trans content, with the highest precision, recall, and thus F1 score out of all the models for this category.  In particular, Pro-Trans recall has almost tripled from the zero shot approach.  This model also has the highest precision for Neutral content, though coming at the cost of the lowest recall for this category. In terms of Anti-Trans Performance, this model has the highest recall, but the lowest precision.  Ultimately, this model is much more likely to label content as either Pro-Trans or Anti-Trans as opposed to Neutral. We see in Table \ref{tab:subrecall} that this model has the highest recall for all Pro-Trans and Anti-Trans sublabels. 
\begin{table*}
  \caption{Classification Performance Metrics and Recall by Sublabel}
  \label{tab:results}
  \begin{tabular}{lcccccccccc}
    \toprule
    & \multicolumn{3}{c}{Anti-Trans (n=52)} & \multicolumn{3}{c}{Pro-Trans (n=160)} & \multicolumn{3}{c}{Neutral (n=88)} \\
    \cmidrule(l){2-4}\cmidrule(l){5-7}\cmidrule(l){8-10}
 {\textbf{Model}}&{P}&{R}&{F1}&{P}&{R}&{F1}&{P}&{R}&{F1}&{Accuracy}\\
    \midrule
    \texttt{LLaMA3}&\textbf{0.43}&0.62&\textbf{0.51}&0.86&0.28&0.42& 0.37&\textbf{0.74}&0.49&0.47\\
    \texttt{LLaMA3+RAG Samples}&0.39&0.58&0.47&0.88&0.58&0.70&0.51&0.68&\textbf{0.58}&0.61\\
    \texttt{LLaMA3+RAG Samples+RAG Taxonomy}&0.37&\textbf{0.67}&0.48&\textbf{0.91}&\textbf{0.76}&\textbf{0.83}&\textbf{0.62}&0.51&0.56&\textbf{0.67}\\
    \bottomrule
  \end{tabular}


  \label{tab:subrecall}
  \begin{tabular}{lccccccccc}
    \toprule
    & \multicolumn{6}{c}{Anti-Trans (n=52)} & \multicolumn{3}{c}{Pro-Trans (n=160)} \\
    \cmidrule(l){2-7}\cmidrule(l){8-10}
 {\textbf{Model}}&{TM}&{ATM}&{XOR}&{TERF}&{RW}&{INTRA}&{CEL}&{REF}&{CON}\\
    \midrule
    \texttt{LLaMA3}&0.54&0.53&\textbf{0.60}&\textbf{0.80}&0.65&0.50&0.38&0.44&0.57\\
    \texttt{LLaMA3+RAG Samples}&0.61&0.41&0.40&0.60&0.59&\textbf{1.00}&0.61&0.67&\textbf{0.71}\\
    \texttt{LLaMA3+RAG Samples+RAG Taxonomy}&\textbf{0.68}&\textbf{0.65}&\textbf{0.60}&\textbf{0.80}&\textbf{0.71}&\textbf{1.00}&\textbf{0.80}&\textbf{0.72}&\textbf{0.71}\\
    \midrule
    Proportion of Sample&0.12&0.07&0.02&0.02&0.15&0.01&0.42&0.16&0.03\\
    \bottomrule
  \end{tabular}
\end{table*}

\subsection{Network Results}
\begin{figure*}
    \centering  
    \begin{subfigure}[b]{0.33\textwidth}\includegraphics[height=0.7\textwidth]{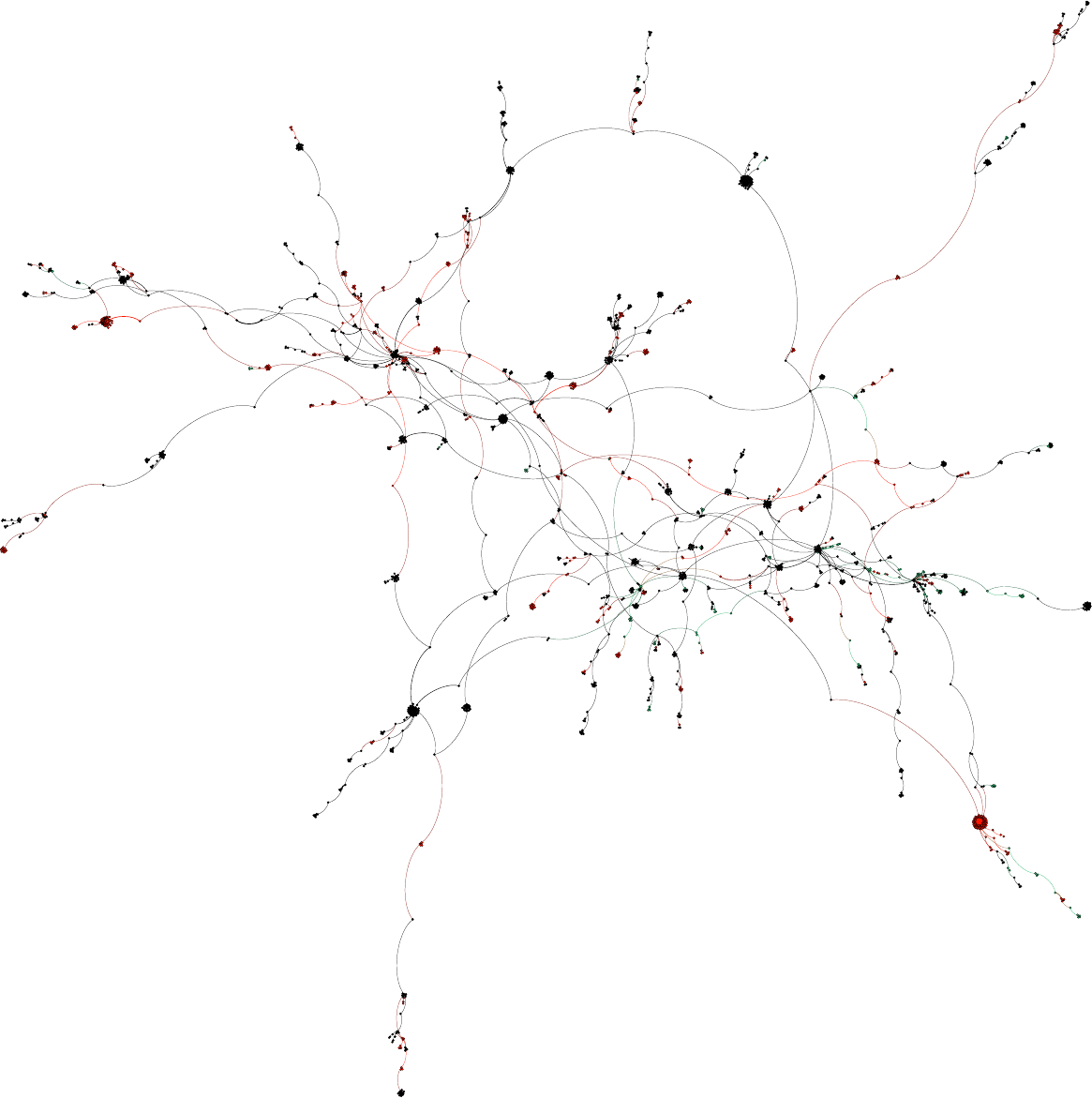}
    \caption{Tag/Reply Network}
    \label{fig:TR}
    \end{subfigure}
    \begin{subfigure}[b]{0.33\textwidth}\includegraphics[height=0.7\textwidth]{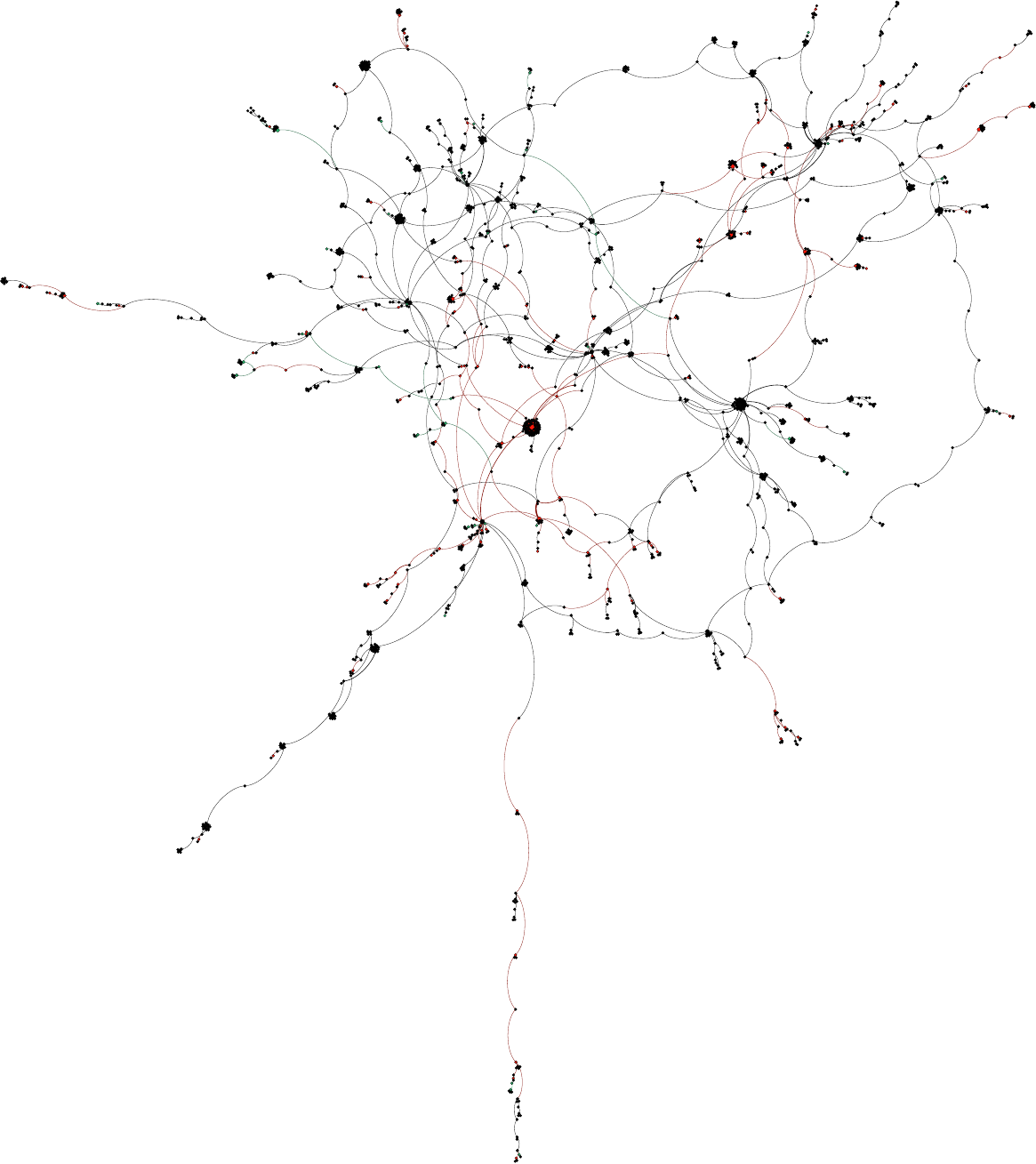}
    \caption{Duet/Stitch Network}
    \label{fig:DS}
    \end{subfigure}
    \caption{Labeled Networks (anti-trans nodes/edges in red, pro-trans in green, neutral in black}
    \label{fig:TRDS}
    \vspace{-1em}
\end{figure*}
We gathered the largest connected component in both the tag/reply \& duet/stitch reply graphs, and ran our LLaMA3 + RAG Examples + RAG Taxonomy classifier on all samples wherein. Looking at the resulting labeled networks in Figure \ref{fig:TRDS}, we see mostly neutral and anti-trans users, with small pockets of pro-trans users.  Anti-trans users outnumber pro-trans users about 5 to 1 in tags/replies, and 2.5 to 1 in duets/stitches.  Assortativity was incredibly low, especially for replies (duet$=-0.51$, reply$=-0.79$, tagged$=-0.49$, stitch=$-0.53$).  When neutral nodes are ignored, this measure is even lower (duet$=-0.64$, reply$=-0.93$, tagged$=-0.62$, stitch=$-0.92$).   

\section{Discussion}

\subsection{Classification}
The baseline version of LLaMA3 was much more likely to label content as Neutral than either Pro-Trans or Anti-Trans, potentially exemplifying cis-normativity and how it both sensationalizes and invisibilizes trans existence \cite{schmitt2019understanding, inmon2023imagining}.  Looking at the particulars, LLaMA3's struggle to recognize \textit{Celebration of Trans Existence} via low recall (.38) demonstrates a common research finding where LLMs struggle with understanding gender-queer in-group language \cite{dorn_harmful_2024}. The off-the-shelf model also fails to find Anti-Trans content invoking \textit{Anti-Transmasculinity} (.53) and \textit{Transmisogyny} (.54). This squares with the ways in which obscuring how transmasculine people are specifically targeted is incentivized in our society \cite{nsambu_za_suekama_racial-class_2023} and how LLM's can struggle to recognize the transmisogyny often implied by anti-transmasculine or exorsexist content, or the transmisogyny inherent to racist degendering of cis women of color \cite{nsambu_za_suekama_racial-class_2023, krell2017transmisogyny, hsu2022irreducible}. The differences in recall for \textit{TERF} and \textit{Right Wing} content indicate furtive research directions into what features are most prominent in these LLM's model of trans-antagonistic actors. 

The \texttt{LLamA3 + RAG Samples} model vastly improved precision on Neutral content and performance overall on all content-- only on subclasses underrepresented in the instances (proportion $\leq$ .07) does recall decrease. However, the performance of \texttt{LLaMA3 + RAG Samples + RAG Taxonomy} yields a meteoric rise in recall for the \textit{Celebration of Trans Existence} as well as all Anti-Trans sublabels.
The improvements yielded by even providing a few samples as context to the classification task points to how strategies for visibility utilized by trans/non-binary creators may result in samples of their content being more informative to our model \cite{nadia_karizat_algorithmic_2021, ellen_simpson_for_2021}. 
Additionally, our codebook in concert with the provided samples being such effective context in identifying Pro-Trans content points to it being fairly easily distinguishable from Neutral and Anti-Trans content when characterizing such content accurately is prioritized.  
The meteoric rise in recall for the \textit{Celebration of Trans Existence} sublabel compared to the base model indicates a schism in perspective between trans/nonbinary community domain experts and the LLM's general training data. 
In queer community, ``Our existence is resistance'' is a resonant slogan, as being openly ourselves defies societal norms \cite{khushboo_sharma_tactics_2021}. 
Incorporating our taxonomy more accurately reflects that reality.  
Further, the improved recall for subgroup sublabels may indicate that our codebook is ameliorating potential biases in our RAG Sample system, and doing so in a way that is much more effective in recalling \textit{Right Wing} content. 

Admittedly, this comes at the cost of a  larger false positive rate of Anti-Trans classification, namely leading to more Neutral content being flagged as Anti-Trans. It is possible that our model is struggling when particular phrases are used in both trans related and irrelevant context (i.e. ``\#protectthechildren'' being used both as an anti-trans dog whistle and as a way to organize against child sexual assault).  Another possibility lies in the accuracy of our annotations, as even with domain experts given a codebook to guide their annotation, trans/nonbinary people can still experience a ``double consciousness'' in navigating this task while aware of how cisgender peers are viewing them \cite{du2006double, wallace2002out}. This points to a need for further research in delineating Anti-Trans from Neutral content.      
\subsection{Network}
It is clear from Figure \ref{fig:TRDS} that anti-trans actors are not only prevalent on the platform, but deeply embedded with ``neutral" users.  This combined with the low assortativity scores, along with visual inspection of the graphs, shows anti-trans actors attacking pro-trans users more often that engaging amongst themselves.  Pro-trans users conversely tended to be either very isolated or found in tight knit clusters, possibly as a defense tactic.

\section{Limitations \& Future Work}
\paragraph{Positionality Statement.} Two of the authors of this work are nonbinary, with one of them being transfeminine.  The remaining authors are cisgender.  
\paragraph{Datasheet.} We host the datasheet for our dataset online \footnote{\nolinkurl{https://github.com/maxynrl/TransTikTokDatasheet}}. Information found there includes ethical considerations in the current and planned future selection and hiring of data annotators, along with our complete set of searched hashtags and our codebook.
\paragraph{Finite Taxonomy.} As much as we've aimed to ground our taxonomy in queer and transfeminist theory, our classification system still fails to capture important nuances in trans/non-binary experience. We look forward to challenging, refining, and extending this taxonomy as we continue to pursue this research topic.
\paragraph{Annotator Choice.} Moving forward, we aim to hire more data annotators in accordance with the ethical standards outlines in our datasheet, particularly trans people of color with varying experiences of transness.  As we do this, it is likely that we will discover biases and limitations in what our current annotations were able to accurately capture about the TikTok data collected.
\paragraph{Trans Scholarship.} As we engage with trans and queer studies, we find that whiteness guides what finds prominence in these fields, inherently making these fields of study racist  \cite{ellison2017we}.  
\paragraph{Speech-to-Text Model.} In future work, we'd like to see how results shift if we try different speech-to-text models for our automated transcription. In particular, we suspect we might see improved classification performance if we utilize a transcription model that is capable of recognizing when the speaker has changed, so as to better understand the spoken content of stitches and duets.
\paragraph{Classification}
We would like to further refine the accuracy of our classifier in future works, as well as introduce classifiers for all sublabels in our taxonomy.  We also plan to explore whether adding more information, such as comments or local network structure, to our inputs will enable more robust classification.
\paragraph{Network Analysis}
Finally, we hope to perform more in-depth network analysis in future works. Particularly, we want to assess how the network patterns we observed here play out with respect to different sublabels of Pro-Trans and Anti-Trans sentiment. Ultimately, we aim to characterize the temporal patterns of network structure characteristics between Anti-Trans and Pro-Trans actors and communities.

\bibliographystyle{ACM-Reference-Format}
\bibliography{antitransharms}

\end{document}